\documentclass[
reprint,
superscriptaddress,
 amsmath,amssymb,
 prc,
]{revtex4-2}
\usepackage{graphicx}
\usepackage{dcolumn}
\usepackage{bm}
\usepackage{hyperref}
\usepackage[mathlines]{lineno}

\begin{document}

\preprint{APS/123-QED}

\title{Practical Considerations for Measuring Global Spin Density Matrix Elements of Vector Mesons in Heavy-Ion Collisions}

\author{Gavin Wilks}
\affiliation{University of Illinois at Chicago, Chicago, Illinois 60607, USA}
\author{Xu Sun}
\affiliation{The Institute of Modern Physics, Chinese Academy of Sciences, Lanzhou 730000, China}
\affiliation{School of Nuclear Science and Technology, University of Chinese Academy of Sciences, Beijing 100049, China}
\affiliation{State Key Laboratory of Heavy Ion Science and Technology, Institute of Modern Physics, Chinese Academy of Sciences, Lanzhou 730000, China}
\author{Zhenyu Ye}
\affiliation{Lawrence Berkeley National Laboratory, Berkeley, CA 94720}






\date{\today}

\begin{abstract}
The STAR Collaboration has reported a significant $\phi$-meson global spin alignment ($\rho_{00}$) signal in Au+Au collisions at $\sqrt{s_{NN}}\leq62$ GeV by measuring the polar angle distribution of $\phi$-meson daughters with respect to the orbital angular momentum (OAM) direction of the collision system. In this paper, a new method is explored for studying vector-meson global spin alignment in heavy-ion collisions by examining the two dimensional polar and azimuthal angle distribution. This method allows simultaneous extraction of $\rho_{00}$ and off-diagonal spin density matrix elements (SDMEs), providing unique access to local quark-antiquark spin correlations and spin hydrodynamics in quark-gluon plasma (QGP). The new 2D method also removes potential biases from non-zero off-diagonal SDMEs on $\rho_{00}$ with the 1D method. A detailed procedure to correct for detector acceptance and resolution effects is also presented and validated by simulation studies. 
\end{abstract}

\maketitle


\section{\label{sec:intro}Introduction}
Since the discovery of a positive $\phi$-meson global spin alignment ($\rho_{00}>1/3$) in Au+Au collisions at $\sqrt{s_{NN}}\leq62$ GeV~\cite{Nature,NST35}, significant effort has been put towards understanding the origin of the signal. It is challenging by conventional mechanisms~\cite{PLB629,Magnetic,Helicity,Axial,Vorticity} to simultaneously describe the observed $\phi$-meson $\rho_{00}$ and the global spin polarization $P_{H}$ of $\Lambda$ baryons \cite{Nature2017,PhysRevC.98.014910,PhysRevC.104.L061901,PhysRevC.108.014910}. Several new models have been proposed to accommodate these signals seen by the STAR Collaboration. One model postulates the existence of a $\phi$-meson strong force field, where fluctuations in this field would induce global spin alignment~\cite{Sheng1,Sheng2,Sheng3,Sheng4,Sheng5}. Another model, which utilizes gauge/gravity duality, suggests that a non-zero helicity-frame spin alignment induced by the motion of $s\bar{s}$ pairs relative to the thermal background in heavy-ion collisions can generate a global spin alignment signal compatible with the data~\cite{HelicityFrame}. A more recent model introduces local strange-quark-antiquark spin correlations to describe the $\phi$-meson $\rho_{00}$ and $\Lambda$-baryon $P_{H}$ signals \cite{PhysRevD.109.114003}. This model also predicts nonzero off-diagonal spin density matrix elements (SDMEs). Therefore, it is very interesting to experimentally measure the off-diagonal SDMEs and compare them with the model predictions. 

The $\phi$-meson $\rho_{00}$ results from the STAR Collaboration utilized a 1-dimensional (1D) approach, where only the polar angular ($\theta^{\ast}$) dimension was considered~\cite{Nature,NST35}. When integrating over the azimuthal angle $\beta$, it was assumed that the off-diagonal SDMEs are zero. Biases from off-diagonal SDMEs to extracted $\rho_{00}$ in the presence of event plane smearing have been discussed in Ref.~\cite{Aihong}. 

In addition to the possible impact of off-diagonal SDMEs on the extraction of $\rho_{00}$, these elements can influence observables of the chiral magnetic effect (CME). Specifically, the spin coherence parameter $\text{Re}\left(\rho_{1-1}\right)$ for the $\rho$ meson has been shown to have an effect on the CME observables related to the $\Delta\gamma_{112}$ correlator, the $R_{\Psi_{2}}\left(\Delta S\right)$ correlator, and the signed balance functions~\cite{CME}. Off-diagonal SDMEs can also directly probe how different components of the angular momentum project onto spin space; therefore, making them crucial for understanding hadronization in vortical environments~\cite{offdiag1,offdiag2,offdiag3}.

In this paper, a method is introduced to simultaneously extract $\rho_{00}$ and off-diagonal SDMEs by studying both polar and azimuthal angular dimensions, eliminating potential biases from off-diagonal SDMEs to $\rho_{00}$. The event plane resolution corrections are analytically solved for all SDME quantities and Monte Carlo simulations are used to confirm the formalism, highlighting the importance of considering event plane smearing in efficiency corrections. The effect of the transverse momentum ($p_{T}$) resolution is also considered, which could bias $\rho_{00}$ and off-diagonal SDMEs when not considered properly in the correction procedure. A detailed procedure to correct for detector acceptance and resolution effects is also presented and validated by simulation studies.

\section{\label{sec:gsa}Global Spin Density Matrix Elements of Vector Mesons}



\begin{figure}[b]
\includegraphics[width=0.4\textwidth]{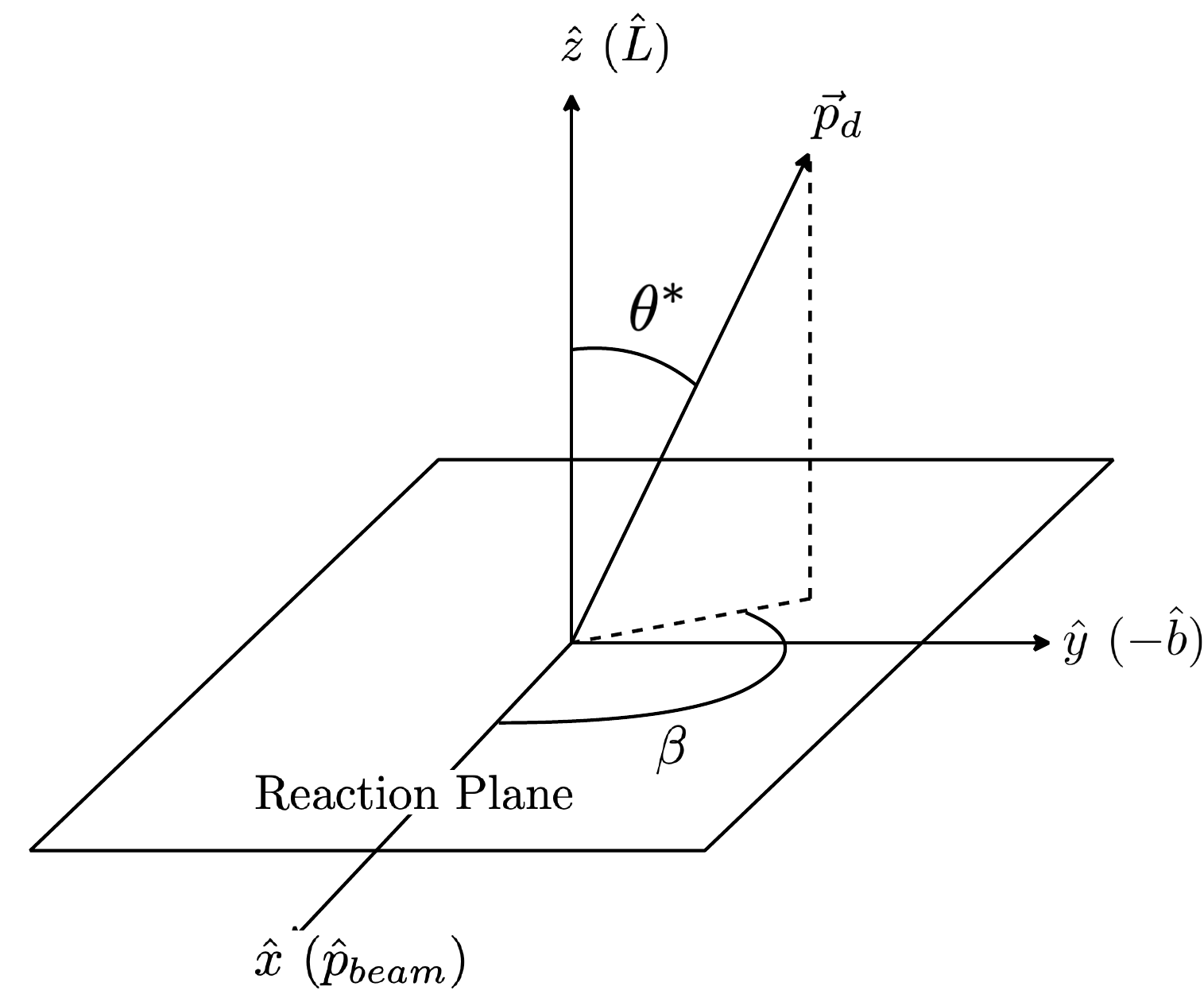}
\caption{\label{fig:coords}Global frame coordinate system: $\vec{p}_{d}$ is the momentum of a daughter particle in the vector meson's rest frame.}
\end{figure}

For spin-1 vector mesons, the probability densities of spin states are captured by the $3\times3$ spin density matrix, $\rho$. The $\rho$ matrix is a Hermitian positive semi-definite matrix with unit trace. The global spin alignment signal is quantified by the diagonal element $\rho_{00}$, which indicates the probability density of vector mesons in the spin-0 state. A deviation of $\rho_{00}$ from $1/3$, indicates a global spin alignment signal. The off diagonal elements $\rho_{mm'}$ $(m\neq m')$, indicate the quantum coherence between the 3 spin states. Considering a two-body decay of a vector meson into two pseudo-scalar mesons, the 2-dimensional angular distribution of a daughter in the vector meson's rest frame is: 
\begin{align}
    \frac{d^{2}N}{d\cos{\theta^{\ast}}d\beta} = 
    &\frac{3}{8\pi}\bigr[\left(1-\rho_{00}\right)+\left(3\rho_{00}-1\right)\cos^{2}{\theta^{\ast}} \nonumber \\
    &-\sqrt{2}\text{Re}\left(\rho_{10}-\rho_{0-1}\right)\sin{2\theta^{\ast}}\cos{\beta} \nonumber \\
    &+\sqrt{2}\text{Im}\left(\rho_{10}-\rho_{0-1}\right)\sin{2\theta^{\ast}}\sin{\beta} \nonumber \\
    &-2\text{Re}\left(\rho_{1-1}\right)\sin^{2}{\theta^{\ast}}\cos{2\beta} \nonumber \\
    &+2\text{Im}\left(\rho_{1-1}\right)\sin^{2}{\theta^{\ast}}\sin{2\beta} \bigr],
    \label{eq:start}
\end{align}
where $\theta^{\ast}$ is the polar angle between the quantization axis (global frame +z-axis, OAM direction) and the daughter's momentum vector, and $\beta$ is the azimuthal angle of the daughter's momentum vector in the reaction plane relative to the beam direction (global frame +x-axis)~\cite{Schilling,Vorticity}. Figure~\ref{fig:coords} depicts the $\theta^{\ast}$ and $\beta$ angles in the global frame. In the global frame, the +x-axis is along the beam direction, the +y-axis is antiparallel to the impact parameter ($\hat{b}$), and +z-axis completes the right-hand coordinate system and is aligned with the orbital angular momentum (OAM) direction. This is different than the lab frame, where the +z-axis is along the beam direction, the +y-axis is orthogonal to the ground, while the +x-axis completes the right hand coordinate system. The azimuthal angle of the reaction plane in the lab frame is known as the angle of the reaction plane, $\Psi_{r}$, where the reaction plane spans the beam direction and $\hat{b}$.

Due to a finite number of particles produced in heavy-ion collisions and a finite detector acceptance, the true angle $\Psi_{r}$ of $\hat{b}$ cannot be known. Therefore, approximation methods described in~\cite{flow5} are used to estimate $\Psi_{r}$. The particle emission angle relative to the reaction plane can be written as a Fourier series:
\begin{equation}
    \frac{dN}{d\phi}=\frac{1}{2\pi}\left(1+\sum_{m=1}^{\infty}2v_{m}\cos\left[m\left[(\phi-\Psi_{m}\right)\right]\right) \label{eq:fourier}.
\end{equation}

Through this decomposition, the angle of the reaction plane can be estimated for each harmonic order, $m$, known as the event plane angle, $\Psi_{m}$~\cite{flow3}. 

When estimating the reaction plane using harmonic event planes, there is an associated resolution, calculated by, 
\begin{equation}
    R_{m}=\langle\cos{\left[m\left(\Psi_{m}-\Psi_{r}\right)\right]}\rangle, \label{eq:res}
\end{equation}
where $\langle\rangle$ represents the average over many events. 
Throughout this paper, the following definition is used: 
\begin{equation}
    \Delta{\Psi_{m}}=\Psi_{m}-\Psi_{r}.
\end{equation}
In data analysis, this resolution can be estimated through the sub-event method, see equations 14 and 15 in~\cite{flow5}. However, in a simulated environment, $\Psi_{r}$ is known and can be artificially smeared to $\Psi_{m}$ by a given resolution through the following probability density function, 
\begin{eqnarray}
    \frac{dP}{d\left(\Delta\Psi_{m}\right)}=\frac{1}{2\pi}\Biggr[&&\exp\left(-\frac{\chi_{m}^{2}}{2}\right)+\sqrt{\frac{\pi}{2}}\chi_{m}\cos{\left(m\Delta\Psi_{m}\right)} \nonumber \\
    &&\times\exp\left(-\frac{\chi_{m}^{2}\sin^{2}{\left(m\Delta\Psi_{m}\right)}}{2}\right) \nonumber \\
    &&\times\left(1+\text{erf}\left(\frac{\chi_{m}\cos{\left(m\Delta\Psi_{m}\right)}}{\sqrt{2}}\right)\right)\Biggr]. \label{eq:eppdf}
\end{eqnarray}
Where $\chi_{m}$ has the following relation to the event plane resolution,
\begin{equation}
    R_{m}=\sqrt{\frac{\pi}{8}}\chi_{m}\exp{\left(-\frac{\chi_{m}^{2}}{4}\right)\left[I_{0}\left(\frac{\chi_{m}^{2}}{4}\right)+I_{1}\left(\frac{\chi_{m}^{2}}{4}\right)\right]}. \label{eq:chi}
\end{equation}
$I_{0}$ and $I_{1}$ are modified Bessel functions of the first kind~\cite{flow3}.

When analyzing real data in the global frame, $\Psi_{m}$ can be used to estimate the direction of $\hat{b}$; therefore, the daughter kaon angles $\theta^{\ast}$ and $\beta$ are also estimations, leading to a shift in the observed SDMEs. The spin density matrix from the event plane, $\rho^{\Psi_{m}}$, can be be related to the spin density matrix from the reaction plane, $\rho$, by considering a spin-1 space rotation of $\rho$ around the global frame +x-axis. The spin-1 x-rotation operator can be written as follows,
\begin{eqnarray}
    \text{R}_{x}\left(\Delta\Psi_{m}\right)&&=e^{-i\Delta\Psi_{m} J_x} \nonumber \\
    &&=\begin{pmatrix}
    \frac{1+\cos{\Delta\Psi_{m}}}{2} & \frac{-i\sin{\Delta\Psi_{m}}}{\sqrt{2}} & \frac{-1+\cos{\Delta\Psi_{m}}}{2}  \\
    \frac{-i\sin{\Delta\Psi_{m}}}{\sqrt{2}} & \cos{\Delta\Psi_{m}} & \frac{-i\sin{\Delta\Psi_{m}}}{\sqrt{2}}  \\
    \frac{-1+\cos{\Delta\Psi_{m}}}{2} & \frac{-i\sin{\Delta\Psi_{m}}}{\sqrt{2}} & \frac{1+\cos{\Delta\Psi_{m}}}{2} 
\end{pmatrix},
\end{eqnarray}
where $J_x$ is the total x-angular momentum operator in the spin-1 basis. 
The rotation operator is applied to $\rho$: 
\begin{equation}
    \rho^{\Psi_{m}}=\text{R}_{x}\left(\Delta\Psi_{m}\right)\rho\text{R}_{x}\left(\Delta\Psi_{m}\right)^{\dagger}.
    \label{eq:epfromrp}
\end{equation}

Using the resulting matrix from Eq.~(\ref{eq:epfromrp}), the parameters in Eq.~(\ref{eq:start}) are calculated in the event plane frame in terms of the reaction plane frame parameters and $\Delta\Psi_{m}$. The relations are as follows: 

\begin{eqnarray}
    \rho_{00}^{\Psi_{m}}=&&\rho_{00}\frac{1+\cos{2\Delta\Psi_{m}}}{2}\nonumber \\ 
    &&+\left(1-\rho_{00}+2\text{Re}\left(\rho_{1-1}\right)\right)\frac{1-\cos{2\Delta\Psi_{m}}}{4} \nonumber\\    
    &&+\frac{\sqrt{2}}{2}\text{Im}\left(\rho_{10}-\rho_{0-1}\right)\sin{2\Delta\Psi_{m}}, \label{eq:rho00rp}
\end{eqnarray}
\begin{eqnarray}
    \text{Re}\left(\rho_{10}^{\Psi_{m}}-\rho_{0-1}^{\Psi_{m}}\right)=&&\text{Re}\left(\rho_{10}-\rho_{0-1}\right)\cos{\Delta\Psi_{m}}\nonumber \\ 
    &&-\sqrt{2}\text{Im}\left(\rho_{1-1}\right)\sin{\Delta\Psi_{m}}, \label{eq:rerp}
\end{eqnarray}
\begin{eqnarray}
    \text{Im}&&\left(\rho_{10}^{\Psi_{m}}-\rho_{0-1}^{\Psi_{m}}\right)=\text{Im}\left(\rho_{10}-\rho_{0-1}\right)\cos{2\Delta\Psi_{m}}\nonumber \\
    &&-\frac{1}{2\sqrt{2}}\left[3\rho_{00}-1-2\text{Re}\left(\rho_{1-1}\right)\right]\sin{2\Delta\Psi_{m}}, \label{eq:imrp}
\end{eqnarray}
\begin{eqnarray}
    \text{Re}\left(\rho_{1-1}^{\Psi_{m}}\right)=&&\frac{1}{8}\bigr[3\rho_{00}-1+6\text{Re}\left(\rho_{1-1}\right)\nonumber \\
    &&-\left(3\rho_{00}-1-2\text{Re}\left(\rho_{1-1}\right)\right)\cos{2\Delta\Psi_{m}}\bigr] \nonumber\\
    &&-\frac{1}{2\sqrt{2}}\text{Im}\left(\rho_{10}-\rho_{0-1}\right)\sin{2\Delta\Psi_{m}}, \label{eq:rerho1n1rp}
\end{eqnarray}
\begin{eqnarray}
    \text{Im}\left(\rho_{1-1}^{\Psi_{m}}\right)=&&\text{Im}\left(\rho_{1-1}\right)\cos{\Delta\Psi_{m}} \nonumber \\
    &&+\frac{1}{\sqrt{2}}\text{Re}\left(\rho_{10}-\rho_{0-1}\right)\sin{\Delta\Psi_{m}}.\label{eq:imrho1n1rp}
\end{eqnarray}
When measuring these quantities in data, the average of each term over many events is extracted. Keeping in mind that the parameters in $\rho$ are independent of the event plane smearing terms involving $\Delta\Psi_{m}$, the average over many events is taken and the reaction plane frame parameters in terms of event plane frame parameters and the averages of the cosine terms are solved. The averages of the sine terms are 0 since this is an average of an odd function over the even distribution in Eq.~(\ref{eq:eppdf}). Ignoring the average brackets for the SDMEs, the following relations are derived:
\begin{eqnarray}
    \rho_{00}=&&\frac{1}{4\langle\cos{2\Delta\Psi_{m}}\rangle}\biggr[\left(3+\langle\cos{2\Delta\Psi_{m}}\rangle\right)\rho_{00}^{\Psi_{m}} \nonumber \\
    &&\left(-1+\langle\cos{2\Delta\Psi_{m}}\rangle\right)\left(1+2\text{Re}\left(\rho_{1-1}^{\Psi_{m}}\right)\right)\biggr], \label{eq:rho00} 
\end{eqnarray}
\begin{equation}
    \text{Re}\left(\rho_{10}-\rho_{0-1}\right)=\frac{\text{Re}\left(\rho_{10}^{\Psi_{m}}-\rho_{0-1}^{\Psi_{m}}\right)}{\langle\cos\Delta\Psi_{m}\rangle}, \label{eq:rerho} \\
\end{equation}
\begin{equation}
    \text{Im}\left(\rho_{10}-\rho_{0-1}\right)=\frac{\text{Im}\left(\rho_{10}^{\Psi_{m}}-\rho_{0-1}^{\Psi_{m}}\right)}{\langle\cos{2\Delta\Psi_{m}}\rangle}, \label{eq:imrho} \\
\end{equation}
\begin{eqnarray}
    \text{Re}\left(\rho_{1-1}\right)=\frac{1}{8}\Biggr[&&\frac{1+2\text{Re}\left(\rho_{1-1}^{\Psi_{m}}\right)-3\rho_{00}^{\Psi_{m}}}{{\langle\cos{2\Delta\Psi_{m}}\rangle}} \nonumber \\
    &&+\left(-1+6\text{Re}\left(\rho_{1-1}^{\Psi_{m}}\right)+3\rho_{00}^{\Psi_{m}}\right)\Biggr],\label{eq:rerho1n1}
\end{eqnarray}
\begin{equation}
    \text{Im}\left(\rho_{1-1}\right)=\frac{\text{Im}\left(\rho_{1-1}^{\Psi_{m}}\right)}{\langle\cos\Delta\Psi_{m}\rangle}. \label{eq:imrho1n1}
\end{equation}
A notable feature of these relations is the dependence of $\rho_{00}$ and $\text{Re}\left(\rho_{1-1}\right)$ on both $\rho_{00}^{\Psi_{m}}$ and $\text{Re}\left(\rho_{1-1}^{\Psi_{m}}\right)$, indicating a coupling of these terms when the event plane resolution is finite. This is a potential downfall for the 1-dimensional method, where an integration over the $\beta$ angle in Eq.~(\ref{eq:start}) is performed, only allowing for the measurement of $\rho_{00}^{\Psi_{m}}$. In this case, a true $\text{Re}\left(\rho_{1-1}\right)$ will lead to a signal of $\rho_{00}^{\Psi_{m}}$, which the 1-dimensional $\cos{\theta^{\ast}}$ method cannot correct for, resulting in a biased final signal of $\rho_{00}$. For $m=1$, the values $\langle\cos{\Delta\Psi_{1}}\rangle$ and $\langle\cos{2\Delta\Psi_{1}}\rangle$ range from 0 to 1, based on how well the true reaction plane is estimated. When $m=2$, $\langle\cos{2\Delta\Psi_{2}}\rangle$ also ranges from 0 to 1, but $\langle\cos{\Delta\Psi_{2}}\rangle=0$, since Eq.~(\ref{eq:eppdf}) is periodic over $\pi$, but $\cos{\Delta\Psi_{2}}$ is periodic over $2\pi$. This prohibits extraction of $\text{Re}\left(\rho_{10}-\rho_{0-1}\right)$ and $\text{Im}\left(\rho_{1-1}\right)$ from the second order event plane in data, as terms extracted relative to the event plane frame angle are forced to 0 by $\langle\cos{\Delta\Psi_{2}}\rangle$. 

\section{\label{sec:simulation}Simulation Studies}

To test the event plane correction formalism in Eqs.~(\ref{eq:rho00}--\ref{eq:imrho1n1}), a Monte Carlo (MC) simulation sample of the decay, $\phi\rightarrow K^{+}+K^{-}$, was created with 500M events. 
The results of this simulation would be qualitatively similar to other two-body decays of vector mesons into pseudo-scalar mesons, such as $K^{\ast}\rightarrow K^{\pm}\pi^{\mp}$ and $\rho\rightarrow\pi^{+}\pi^{-}$.
The model used inputs from Pythia8 simulations or published data for 20-60\% centrality Au+Au collisions. The $\phi$-meson $p_{T}$ distribution was weighted by a Levy fit to the transverse momentum ($p_{T}$) spectrum from~\cite{ptspectrum}. Pythia8 p+p collisions were sampled to estimate the rapidity ($y$) spectrum for $\phi$-mesons~\cite{pythia8}. The $\phi$-mesons were simulated with $1.2<p_{T}<4.2$ GeV/c and $|y|<1$, to match the kinematic selections used by the STAR Collaboration in~\cite{Nature}. The reaction plane angle followed a randomly generated uniform distribution. The input elliptic flow ($v_{2}$), which describes the particle emission relative to the reaction plane, was derived from fitting equation 2 from~\cite{flowfunction} to the $p_{T}$ dependent $v_{2}$ data in~\cite{19p6flow}. By construction, the daughter $K^{+}$ and $K^{-}$ distributions in the $\phi$-meson rest frame were isotropic, which translates to $\rho_{00}=1/3$ and $\text{Re}\left(\rho_{10}-\rho_{0-1}\right)=\text{Im}\left(\rho_{10}-\rho_{0-1}\right)=\text{Re}\left(\rho_{1-1}\right)=\text{Im}\left(\rho_{1-1}\right)=0$. When simulating any deviation of these parameters, Eq.~(\ref{eq:start}) was used to weight the 2-dimensional $\cos\theta^{\ast},\beta$ distribution. 

\begin{figure*}
    \includegraphics[width=\textwidth, trim=13pt 30pt 10pt 10pt, clip]{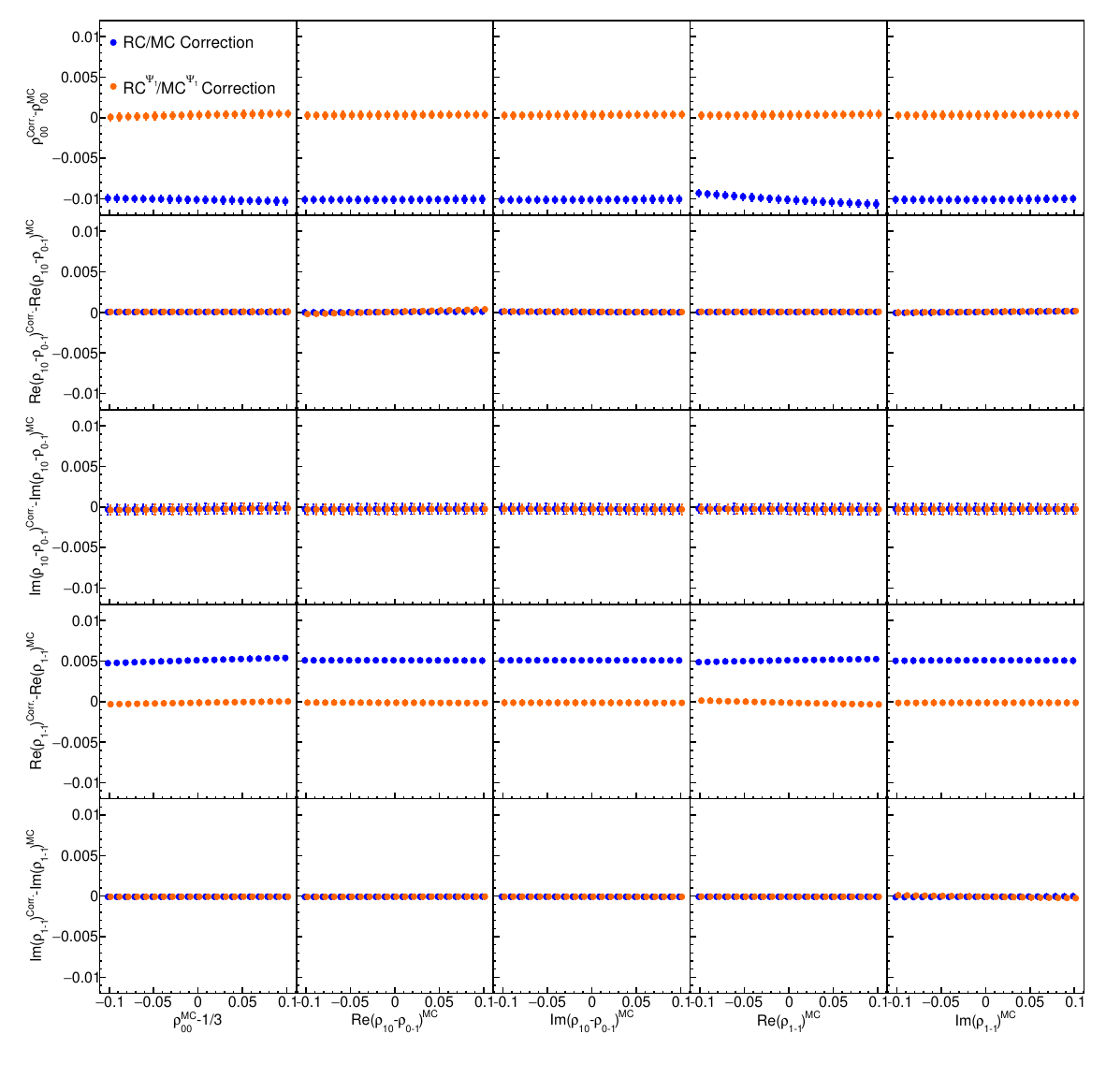}
    \caption{\label{fig:epres}2-dimensional event plane resolution corrected results with and without event plane smearing: Blue points correspond to using the reaction plane yield correction and the orange points correspond to the event plane yield correction.} 
\end{figure*}

With this MC model setup, the ability to extract $\rho$ parameters was tested when event plane smearing is present, following directly from the relations in Eqs.~(\ref{eq:rho00}--\ref{eq:imrho1n1}). Also examined were the effects of including or excluding event plane smearing in simulation when building a $\cos\theta^{\ast}$,$\beta$ yield correction matrix. This study started with the MC generator-level information, listed at the beginning of this section, and then variable reweighting of the $\cos\theta^{\ast}$,$\beta$ distributions was applied to simulate the $\rho$ parameters. The event plane angles were smeared relative to the reaction plane angle following Eqs.~(\ref{eq:eppdf}--\ref{eq:chi}), where $R_{1}=0.6$ to roughly match the event plane resolution in STAR. The following acceptance cuts were applied on the daughter kaons: $|\eta|<1$, $p_{T}>0.1$ GeV/c, and $|\mathbf{p}|<10$ GeV/c. Various detector efficiencies from the STAR experiment were also applied to the daughter kaons leaving reconstructed kaon tracks. The reconstructed $\phi$-mesons distributions were then filled according to the $K^{+}$ momentum angles in the $\phi$-meson rest frame relative to the event plane, $\cos{\theta^{\ast\Psi_{m}}}$ and $\beta^{\Psi_{m}}$, labeled $\text{RC}^{\Psi_{m}}$. In the case where there was no smearing of the event plane for the reconstructed $\phi$-mesons, the reconstructed $\cos\theta^{\ast}$,$\beta$ distributions are labeled $\text{RC}$. Similarly, the MC level $\cos\theta^{\ast}$,$\beta$ and $\cos{\theta^{\ast\Psi_{m}}}$,$\beta^{\Psi_{m}}$ distributions are labeled $\text{MC}$ and $\text{MC}^{\Psi_{m}}$, respectively. In the isotropic case, which is used to extract the $\phi$-meson yield corrections, the labels are changed to $\text{RC}_{0}$, $\text{RC}^{\Psi_{m}}_{0}$, $\text{MC}_{0}$, and $\text{MC}^{\Psi_{m}}_{0}$. For all of these distributions 10 equal sized bins for $\cos{\theta^{\ast}}$ from -1 to 1, and 10 equal sized bins for $\beta$ from 0 to $2\pi$ were used. These bin sizes were chosen to mimic realistic binning for experimental heavy-ion collision data. 

\begin{figure*}
    \includegraphics[width=0.8\textwidth, trim=15pt 5pt 50pt 5pt, clip]{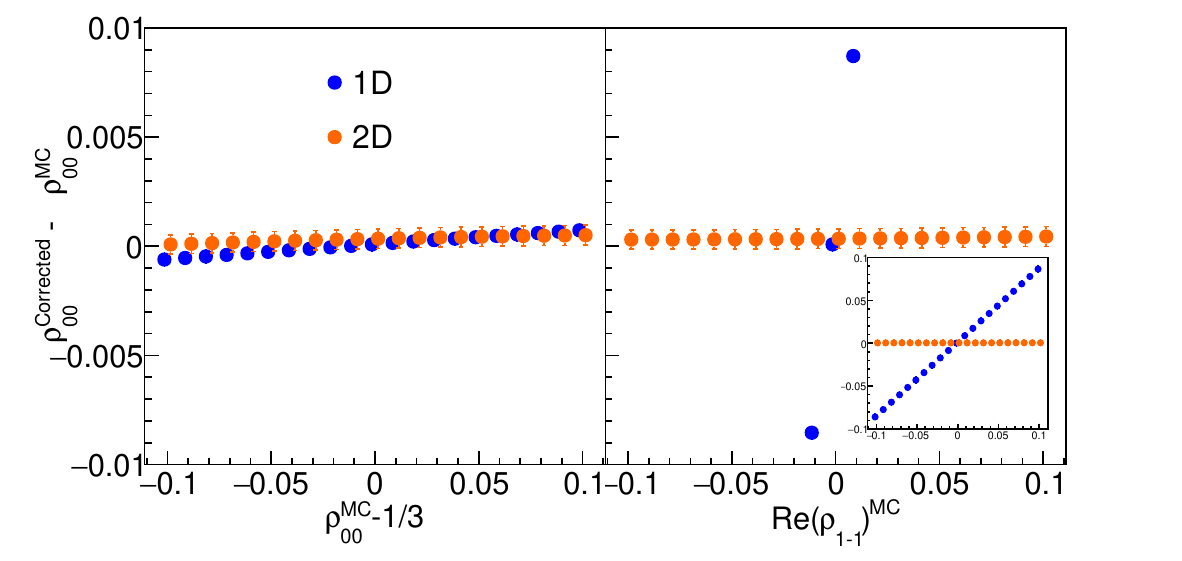}
    \caption{\label{fig:epres1D2D}1-dimensional and 2-dimensional event plane resolution correction: Blue points correspond to using the 1-dimensional method and orange points correspond the 2-dimensional method.}  
\end{figure*}  

The distribution $\text{RC}^{\Psi_{m}}$ was corrected for various $\rho$ parameter inputs by dividing each distribution by the overall efficiency of $\phi$-meson reconstruction in the isotropic case. The correction procedure was tested with the event plane smeared efficiency, $\text{RC}^{\Psi_{m}}_{0}/\text{MC}^{\Psi_{m}}_{0}$, and the reaction plane level efficiency, $\text{RC}_{0}/\text{MC}_{0}$. Note that the isotopic case was used since the true $\rho$ values in data are not known and the corrections are assumed to be independent of the input $\rho$. 
In the previous 1-dimensional method, event plane smearing was ignored in the correction procedure, and any difference between final corrected $\rho$ parameters and the MC level parameters due to this missing feature is quantified here. 
To quantify any difference between final corrected $\rho$ parameters and the MC level parameters due to ignoring the smearing of the event plane, the efficiency yield corrections are applied by dividing $\text{RC}^{\Psi_{m}}$ by $\text{RC}^{\Psi_{m}}_{0}/\text{MC}^{\Psi_{m}}_{0}$ and $\text{RC}_{0}/\text{MC}_{0}$, separately. Then, the event plane frame parameters $\rho^{\Psi_{m}}$ were extracted from the corrected $\cos{\theta^{\ast\Psi_{m}}},\beta^{\Psi_{m}}$ distribution by fitting with the following function:
\begin{eqnarray}
    &&\frac{d^{2}N}{d\cos{\theta^{\ast\Psi_{m}}}d\beta^{\Psi_{m}}} = 
    N_{0}\biggr[1-\rho_{00}^{\Psi_{m}}\nonumber \\
    &&+\left(3\rho_{00}^{\Psi_{m}}-1\right)\cos^{2}{\theta^{\ast\Psi_{m}}} \nonumber \\
    &&-\sqrt{2}\text{Re}\left(\rho_{10}^{\Psi_{m}}-\rho_{0-1}^{\Psi_{m}}\right)\sin{2\theta^{\ast\Psi_{m}}}\cos{\beta^{\Psi_{m}}} \nonumber \\
    &&+\sqrt{2}\text{Im}\left(\rho_{10}^{\Psi_{m}}-\rho_{0-1}^{\Psi_{m}}\right)\sin{2\theta^{\ast\Psi_{m}}}\sin{\beta^{\Psi_{m}}} \nonumber \\
    &&-2\text{Re}\left(\rho_{1-1}^{\Psi_{m}}\right)\sin^{2}{\theta^{\ast\Psi_{m}}}\cos{2\beta^{\Psi_{m}}} \nonumber \\
    &&+2\text{Im}\left(\rho_{1-1}^{\Psi_{m}}\right)\sin^{2}{\theta^{\ast\Psi_{m}}}\sin{2\beta^{\Psi_{m}}} \biggr],
    \label{eq:fit2D}
\end{eqnarray}
where $N_{0}$ is a free normalization parameter. The final $\rho$ parameters are calculated using Eqs.~(\ref{eq:rho00}--\ref{eq:imrho1n1}). In an actual experiment such as STAR, the distribution RC$^{\Psi_{m}}$ would be replaced by the $\phi$-meson yields extracted in each $\cos{\theta^{\ast\Psi_{m}}}$ and $\beta^{\Psi_{m}}$ bin combination. See Extended Data Figures 1a and 1b in~\cite{Nature} for examples of vector meson yield extraction from STAR data.

The same procedure was repeated for the 1-dimensional method, where all distributions were integrated over the $\beta$ dimensions before any corrections were derived or applied. The 1-dimensional $\rho_{00}^{\Psi_{m}}$ was extracted by fitting the corrected $\phi$-meson yields with Eq.~(\ref{eq:fit2D}) integrated over $\beta^{\Psi_{m}}$:
\begin{equation}
    \frac{dN}{d\cos{\theta^{\ast\Psi_{m}}}}=
    N_{0}\left[1-\rho_{00}^{\Psi_{m}}+\left(3\rho_{00}^{\Psi_{m}}-1\right)\cos^{2}{\theta^{\ast\Psi_{m}}}\right], \label{eq:fit1D}
\end{equation}
The reaction plane frame $\rho_{00}$ was then calculated following Eq. (15) derived in Ref.~\cite{Aihong}, rewritten here:
\begin{equation}
    \rho_{00}=\frac{4}{1+3\langle\cos{2\Delta\Psi_{m}}\rangle}\left(\rho_{00}^{\Psi_{m}}-\frac{1}{3}\right)+\frac{1}{3}. \label{eq:1drho}
\end{equation}
This equation can also be derived from Eq.~(\ref{eq:rho00}) by setting all off-diagonal elements to zero, averaging over all events, and then solving for $\rho_{00}$.

\begin{figure*}
    \includegraphics[width=\textwidth, trim=13pt 30pt 10pt 10pt, clip]{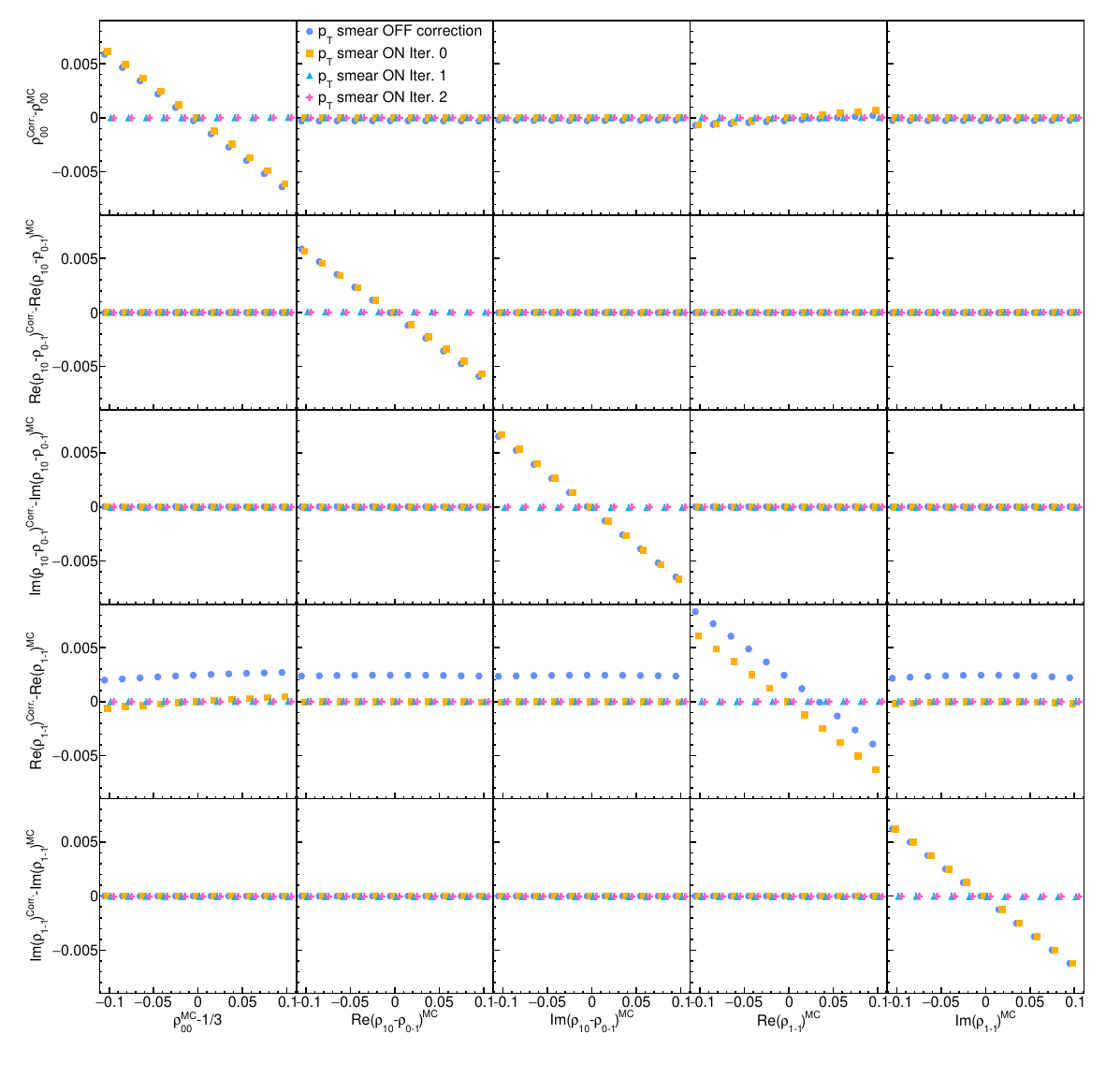}
    \caption{ \label{fig:ptres}$p_{T}$ resolution correction with iterative method: The blue circles correspond to $p_{T}$ smearing off in the correction procedure. Yellow squares correspond to the first attempt (iteration 0) at correcting the $p_{T}$ resolution effect using isotropic $\cos{\theta^{\ast}},\beta$ MC. The cyan triangles and pink crosses correspond to the first and second iteration of input MC $\rho$ parameters, respectively.} 
\end{figure*}

In Figure.~\ref{fig:epres}, it is shown that ignoring the smearing of the event plane in the correction procedure can lead to systematic shifts in the values of the SDME parameters. This figure also shows the ability to correct for the event plane resolution for all SDME parameters regardless of which truth level parameter was varied, confirming the validity of Eqs.~(\ref{eq:rho00}--\ref{eq:imrho1n1}). Figure.~\ref{fig:epres1D2D} shows the comparison of $\rho_{00}$ extraction with the 1-dimensional and 2-dimensional extraction methods with variable input $\rho_{00}$ and $\text{Re}\left(\rho_{1-1}\right)$. The 1-dimesional $\rho_{00}$ was corrected for event plane resolution using Eq.~(\ref{eq:1drho}). It was found that when the true $\rho_{00}$ is varied, both methods could recover the true value within 1\%, with the 2-dimensional method having larger uncertainty and slightly better agreement with the true value. When the true $\text{Re}\left(\rho_{1-1}\right)$ is non-zero, the 2-dimensional correction method can extract the true $\rho_{00}$ value, but the 1-dimensional method cannot, as expected from \eqref{eq:rho00}.

Another important consideration when extracting vector meson SDMEs from particle detector data is $p_{T}$ resolution. 
Since a detector must have some finite precision in particle tracking, it is important to understand how this affects the measured results. 
When measuring distributions relative to $\cos(\theta^{\ast})$ and $\beta$, these angular quantities are directly related to the momenta of the daughter kaons and the $\phi$-meson, and are therefore affected by a finite $p_{T}$ resolution in a non-trivial manner.

Starting with the MC sample described at the beginning of this section, a Gaussian smearing of kaon $p_{T}$ based on a 10\% $p_{T}$ resolution was applied. This value was chosen to emphasize the effect of the $p_{T}$ resolution, and most heavy-ion collisions detectors have $p_{T}$ resolutions on the order of 2-3\% percent or smaller.
To study this effect exclusively, no cuts or event plane smearing were applied, and the final $p_{T}$ smeared distribution $\cos\theta^{\ast\prime}$,$\beta^{\prime}$ was corrected using the isotropic $\rho$ parameter case. If the $p_{T}$ values in the correction procedure are not smeared, this is the same as simply extracting the $\rho$ parameters directly from the $\cos\theta^{\ast\prime}$,$\beta^{\prime}$ distributions. It is observed that ignoring the $p_{T}$ resolution in correction can have a significant effect, given by the blue circles in Figure ~\ref{fig:ptres}. The correction was also tested by taking the ratio of the $p_{T}$ smeared isotropic $\cos\theta^{\ast\prime}$,$\beta^{\prime}$ distribution over the non-smeared distribution, and then dividing the $\cos\theta^{\ast\prime}$,$\beta^{\prime}$ distribution with various input $\rho$ parameters by this ratio. Ideally, this would correct for the $p_{T}$ smearing and the MC level input parameters would be retrieved; however, as seen in Figure~\ref{fig:ptres}, a difference is observed between the corrected $\rho$ values in the $0^{th}$ iteration and the input MC values, displayed as yellow squares. At $p_{T}$ resolutions of 2-3\%, the deviations of corrected SDME values from MC are roughly 10-20\% of what is observed for 10\% $p_{T}$ resolution.


To address the $\rho$ parameter dependence of the $p_{T}$ resolution effect, the corrections can be recalculated using the output $\rho$ parameters from the previous iteration. This iterative method was tested in Figure~\ref{fig:ptres} and it is shown that after just one iteration the input MC $\rho$ values are recovered. Therefore, when analyzers are measuring $\rho$ parameters, iterating the correction procedure at least one time should provide a more accurate measurement of the true value.

\section{\label{sec:conclusion}Conclusion}
The measurement of vector meson global SDMEs in heavy-ion collisions requires careful consideration of various factors, some of which were studied in this paper. Event plane resolution correction formulas were derived for all five global SDME parameters and were validated using a Monte Carlo simulation study with realistic efficiency and acceptance effects. In deriving these corrections, it was discovered that for even order harmonic event planes, it is not possible to extract $\text{Re}\left(\rho_{10}-\rho_{0-1}\right)$ and $\text{Im}\left(\rho_{1-1}\right)$ as these quantities are forced to zero. Through Monte Carlo studies, it was found that event plane smearing is required in the correction procedure, and without it there is a systematic shift of the extracted SDME parameters. Additionally, it was shown that when event plane smearing is present, a 1-dimensional extraction of $\rho_{00}$ from the $\cos{\theta^{\ast}}$ distribution of $\phi$-mesons with a non-zero $\text{Re}\left(\rho_{1-1}\right)$ will bias the output $\rho_{00}$ value. This emphasizes the importance of performing a 2-dimensional extraction of all SDME parameters. The $p_{T}$ resolution was also considered, where it was discovered that the effect from this resolution depends on the true SDME values. By recalculating the particle yield corrections with SDME values from the previous iteration the results converge with the true input values within three iterations. The methods discussed in this paper are crucial for proper measurements of vector meson global SDME parameters.

\section*{\label{sec:ack}Acknowledgments}
Thank you to A.-H. Tang and D. Shen for many fruitful discussions. 
Gavin Wilks is partially supported  by the U.S. Department of Energy under Contract No. DE-FG02-94ER40865. 
Zhenyu Ye is partially supported by the U.S. Department of Energy under Contract No. DE-AC02-05CH11231.
Xu Sun is partially supported by National Natural Science Foundation of China (NSFC) under Grant No. 12475147.

\nocite{*}

\bibliography{maintext}

\end{document}